\documentclass{appolb}
\usepackage{graphicx}

\begin{document}
\title{Size dependence of the largest distance between random points%
\thanks{Presented at the 2013 Summer Solstice International Conference on Discrete Models of Complex Systems, June 27-29, 2013, Warszawa, Poland}%
}
\author{Ma{\l}gorzata J. Krawczyk, Janusz Malinowski, Krzysztof Ku{\l}akowski
\address{Faculty of Physics and Applied Computer Science, AGH University of Science and Technology, al. Mickiewicza 30, 30-059 Krakow, Poland}
\\
}
\maketitle
\begin{abstract}

A set of $N$ points is chosen randomly in a $D$-dimensional volume $V=a^D$, with periodic boundary conditions. For each point $i$, its distance $d_i$ is found to its nearest
neighbour. Then, the maximal value is found, $d_{max}=max(d_i, i=1,...,N)$. Our numerical calculations indicate, that when the density $N/V$=const, $d_{max}$ scales with the 
linear system size as $d^2_{max}\propto a^\phi$, with $\phi=0.24\pm0.04$ for $D=1,2,3,4$.

\end{abstract}

\PACS{05.40.-a, 05.10.Ln, 05.65.+b, 89.75.-k}
  
\section{Introduction}

In our recent paper \cite{my} the problem of penetration of a given area $V$ by deterministic robots, referred to as ants, has been considered. In that model, ants could exchange 
information on areas they had been visited. New scaling relations have been proposed for the time of penetration against the number of ants $N$ and the system size 
$V=a^2$, where $a$ was the linear size of the system. In particular, the time $T$ when all ants know the whole area has been found to be proportional 
to $V^\beta \rho^\delta$, where $\rho=N/V$ was the ant density. The related exponents $\beta$ and $\delta$ have been found numerically as 0.69 and -0.4, respectively. \\

The aim of the present note is to verify the role of the spatial fluctuations of the initial positions of ants in the value of $\beta$. Our question is as follows: provided
that $\rho$ = const, the size dependence of $T$ can be partially attributed to the fact that some ants are initially more far from their neighbours. Then, those ants need more time 
to meet other ants and collect the information from them. How, then, does the maximal distance $d_{max}$ between nearest neighbours depend on the area $V$? \\

In the picture based on diffusion, an ant can be represented by a sphere of radius $r$, and $r^2$ increases linearly with time. Out of this sphere, the probability of meeting of the 
ant is zero. A meeting of two ants is then equivalent with a collision of two spheres. The time $T$ is expected to  be dominated by $d^2_{max}$. This picture is justified for $D=1$ and 
perhaps for $D=2$, because in a low-dimensional space trajectories fill the spheres densely, but not for $D>2$. As we demonstrate below, the obtained scaling relation can be written
in a compact form, the same for $D\le2$ and $D>2$. The price paid for this simplicity is a substitution of the difficult random walk problem by the diffusion model. \\

We note that the relation $d^2_{max}(V)$ can be formulated in terms of the probability distribution of an extremal value \cite {gumbel} of the distance between random points \cite{lord}. 
This formulation should allow to get the mean of the maximal distance to a nearest neighbours through the formalism of generating functions. Yet, the numerical procedure applied here 
seems much simpler than the approach with use of special functions \cite{lord}.\\

In the next section, we describe the simulation developed to get the relation $d_{max}(V)$, and the numerical results for the dimensionality $D$ = 1, 2, 3 and 4. Last section is devoted 
to discussion.\\

\section{Simulation}

The simulation is performed as follows. A $D$-dimensional area $V=a^D$ with periodic boundary conditions is filled with $N=\rho V$ points of randomly selected positions. For each point 
$i$ = 1,...,$N$ we find its distance $d_i$ to its nearest neighbour. Next, we find the maximal value of $d_i$, denoted as $d_{max}$.This is done for $a$ varying with $\rho$ = constant, 
and for $D$= 1, 2, 3 and 4. The number of points is taken as $N=100a^D$.\\

The results are shown in Fig. 1. The adopted density is $N/V$ is... Yet, as we see, the obtained plot indicates the scaling relation $d^2_{max}\propto a^\phi$, with $\phi$ close to 
0.24 for all investigated dimensionalities $D$. The accuracy of $\phi$ is bounded by fitting error which varies from 0.02 to 0.04 for different dimensionalities.\\

\begin{figure}[htb]
\centerline{%
\includegraphics[width=12.5cm]{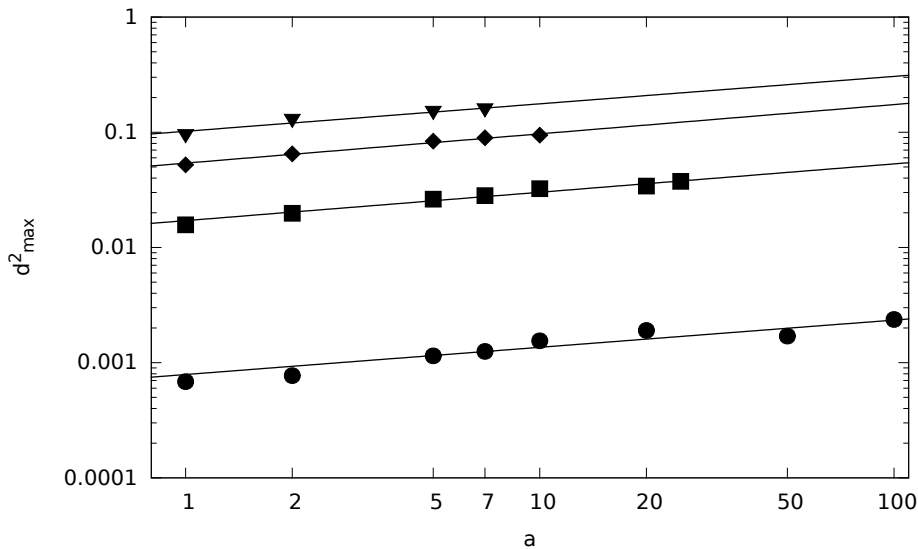}}
\caption{The relation between $d^2_{max}$ and the linear size $a$ of the system, for the dimensionality $D$ = 1, 2, 3, 4 from bottom to top. Each point is an average over 10 runs.}
\label{Fig:F2H}
\end{figure}

\section{Discussion}

To compare the obtained scaling relation with the result $T\propto V^\beta$ for $D=2$ \cite{my}, we substitute $a=V^{1/2}$. This gives $d^2_{max}\propto V^{0.12}$. Provided 
that the sphere area $r^2\propto t$, as follows from the diffusion model, we get the time $T\propto a^{\phi}=V^{0.12}$. This exponent is clearly less than $\beta=0.69$, obtained
in \cite{my}. As the ants are designed to omit their previous paths \cite{my}, an alternative assumption could be done according to the Self-Avoiding-Walk model \cite{flory}, 
where for $D=2$ the rule is $r^2\propto t^{3/2}$. Then we get 
$T\propto (r^2)^{2/3}\propto a^{2\phi /3}=V^{\phi/3}=V^{0.08}$, what is even more far from $V^\beta$. This means that the observed value of the exponent $\beta$ cannot be attributed 
solely to the fluctuations of the initial density of ants. For completeness we may add that $d^2_{max}\propto \rho^{-2/D}$ for purely geometrical reasons. \\

The scaling relation $d^2_{max}\propto a^\phi$ , which appeared as a by-product of the considerations of Search-And-Rescue robots in a labyrinth \cite{my}, can find applications 
also in other areas. An example is a spatial set of sensors; two sensors interact if their distance is not larger than a prescribed value \cite{kucuk,gauger,natka}. Yet, the distance 
in other than geometrical, high-dimensional space can play the same role. Namely, in a network of of time series of biological or economic series of data \cite{kwa} as points 
and their mutual correlations as bonds, a least correlated signal in a network is equivalent to the maximally distant point. There, the signal length is a counterpart of the 
space dimensionality $D$, and the range of the signal -- of the linear system size $a$. With this interpretation, our results can be helpful to identify outliers in data.


\begin{thebibliography}{80}
\bibitem{my} J. Malinowski, J. W. Kantelhardt and K. Ku{\l}akowski, {\it Deterministic ants in labyrinth - information gained by map sharing}, Int. J.
Modern Physics C 24 (2013) 1350035.
\bibitem{gumbel} E. J. Gumbel, {\it Statistics of Extremes}, Columbia UP, New York 1958.
\bibitem{lord} R. D. Lord, {\it The distribution of distance in a hypersphere}, Ann. Math. Statist. 24 (1954) 794.
\bibitem{flory} M. Doi and S. F. Edwards, {\it The Theory of Polymer Dynamics}, Clarendon Press, Oxford 1986. 
\bibitem{kucuk} W. Jia, Y. Fu and J. Wang, {\it Analysis of connectivity for sensor networks using geometrical probability}, Lect. Notes Comp. Sci. 3207 (2004) 601.
\bibitem{gauger} M. Gauger, {\it Integration of Wireless Sensor Networks in Pervasive Computer Scenarios}, Logos Verlag Berlin GmbH 2010.
\bibitem{natka} J. Natkaniec and K. Ku{\l}akowski, {\it When the spatial networks split?}, Lect. Notes Comp. Sci. 5102 (2008) 545.
\bibitem{kwa} J. Kwapie\'n and S. Dro\.zd\.z, {\it Physical approach to complex systems}, Phys. Rep. 515 (2012) 115.
\end{thebibliography}
\end{document}